\documentclass{appolb}
\usepackage{graphicx}
\begin{document}
\title{Facets of Neutrino-Nucleus Interactions%
\thanks{Presented at the XXXV Mazurian Lakes Conference on Physics, Piaski, Poland, September
3-9, 2017}%
}
\author{A.B. Balantekin
\address{University of Wisconsin, Department of Physics, Madison, Wisconsin 53706 USA \linebreak}\\ 
}

%for multiple authors (>5) please use instead
%\author{ AuthorA$^{a}$, AuthorB$^{b}$, AuthorC$^{c}$, AuthorD$^{d}$,AuthorE$^{e}$, AuthorF$^{f}$
%\address{$^a$ affiliationA\\ $^b$ affiliationB\\ $^c$ affiliationC\\ $^d$ affiliationD\\
%$^e$ affiliationE\\ $^f$ affiliationF}} 

\maketitle
\begin{abstract}
Different approaches to the calculation of neutrino-nucleus cross sections are summarized. Potential impact of improving the nuclear physics input into neutrino interactions 
and cross section calculations on uncovering new physics is discussed using the example of reactor anomaly. Importance of a thorough understanding of neutrino interactions in astrophysics and cosmology 
is highlighted. 
\end{abstract}
  
\section{Introduction}
Neutrino physics is closely connected to nuclear physics, a connection which goes beyond  the evident connection between neutrino detection and the nuclear structure of the target isotopes. For example in a core-collapse supernova understanding neutrino cooling of the newly formed proto-neutron star benefits from knowledge of the nuclear equation of state. 
In such environments or in merging of two neutron stars, neutrinos determine the neutron-to-proton ratio, the parameter controlling yields of nucleosynthesis. An old problem in nuclear physics is to accurately calculate neutrino-nucleus cross sections and beta decay rates. A firm knowledge of the nuclear matrix elements for the neutrinoless double beta decay is crucial to assess the experimental outlook for observing possible violation of lepton number, a fundamental symmetry of the Universe. For many aspects of supernova physics we need to know what happens when a 10 to 40 MeV neutrino hits a  nucleus. Longstanding questions include distribution of the Gamow-Teller and tensor strengths as well as the value of the effective axial vector strength factor, $g_A$. As the incoming neutrino energy increases, the contribution of hard to calculate expectation values  increase, including first- and even second-forbidden transitions. Forbidden transitions may be the key to understand decays of isotopes in the nuclear fuel of power reactors and the resulting reactor neutrino spectra. 

Several recent experiments emphasize the need for better nuclear data in connection with fundamental science, either exploring new physics beyond the Standard Model or exploring astrophysical phenomena. For example short-baseline reactor neutrino experiments successfully measured the neutrino parameters they set out to measure, but they also identified an excess of reactor antineutrinos with energies around 5 MeV as well as a reduction from the predicted value of the flux \cite{An:2016srz,An:2015nua,Abe:2015rcp,RENO:2015ksa}. 
This result raises some very interesting nuclear physics questions regarding neutrino interactions some of which we discuss below. 

A key development during the last few decades has been the appreciation of the close relationship between neutrinos and nucleosynthesis as physicists and astronomers ascertained the fact that neutrino properties figure prominently in many astrophysical environments.
Consequently all the properties of neutrinos could significantly impact description of astrophysical environments. Understanding where and how various nuclei are synthesized during the evolution of the Universe is one of the key questions of modern science. 
Element synthesis is thought to be a multi-site and multi-epoch process. 
Tackling the question of the origin of elements requires a multitude of tools: High-quality observations of stellar spectra, laboratory atomic physics data, modeling stellar photospheres as well as theoretical and experimental investigations of the relevant nuclear processes. Typically copious amounts of neutrinos are present in most nucleosynthesis sites. This feature makes neutrino physics and neutrino-nucleus interactions salient components of many nucleosynthesis scenarios.  
The
interaction of the neutrinos with ordinary matter is rather feeble
except when the density is very large. Consequently neutrinos can easily
transfer a significant amount of energy and entropy over astronomical distances. 
(For example almost the entire gravitational binding energy of a pre-supernova
star is released as neutrinos). Clearly such energy transfers could be very
important in astrophysics and cosmology, making a thorough understanding of neutrino interactions crucial to explore many such phenomena. 

Status and challenges of neutrino-nucleus scattering for a wide range of energies was recently summarized in Ref. \cite{Alvarez-Ruso:2017oui}. In this proceedings contribution the discussion is limited to a few examples of interactions of neutrinos with very low energies (up to few tens of MeV) and nuclei. 

\section{Some cross section calculations}

In this section calculations of three different neutrino-nucleus cross sections, chosen to illustrate three different techniques utilized to calculate such cross sections, are briefly discussed. 

Determining the interaction between two nucleons is a long-standing problem. 
During the last decade both the nuclear structure physics and nuclear reactions communities increasingly made use of the effective field theory approach. 
With the advent of the effective field theory methods there had been a renewed 
interest in deriving the nucleon-nucleon interaction from the fundamental theory. In effective field theories describing low-energy physics one integrates over the degrees of freedom associated with physics coming into play at higher energies. However one has to introduce counter terms to cancel divergences which may arise at higher orders. At energies below the pion threshold nucleon-nucleon interaction is particularly simple: $^3S_1 \rightarrow ^3S_0$ transition dominates and one has to introduce a single counter term, dubbed $L_{1A}$, 
characterizing the unknown isovector axial two-body current. The cross sections for the reactions 
\[
\nu_e + d \rightarrow p + p + e^-
\] 
and
\[
\overline{\nu}_e + d \rightarrow n + n + e^+
\] 
can then be calculated in a pionless effective field theory as a function of this unknown term \cite{Butler:1999sv,Butler:2000zp}. The resulting cross sections can be written as 
\[
\sigma (E_{\nu}) = \sigma_0 (E_{\nu}) + L_{1A} \>   \sigma_1 (E_{\nu})
\]
where the terms $\sigma_0 (E_{\nu})$ and $\sigma_1 (E_{\nu})$ can be easily evaluated. The value of $L_{1A}$  can be estimated either from reactor anti-neutrino deuteron breakup reactions \cite{Butler:2002cw} or from solar neutrino experiments \cite{Chen:2002pv,Balantekin:2003ep,Balantekin:2004zj}. From these considerations one obtains a value of 
$L_{1A} \sim 4$ fm$^3$. Very recently this parameter was calculated using lattice QCD  at a renormalization scale set by the physical pion mass to be 
$L_{1A} = 3.9 (0.1) (1.0) (0.3) (0.9) $fm$^3$  \cite{Shanahan:2017bgi}. Hence we have an accurate description of weak breakup of the deuteron and the reverse reaction of proton-proton fusion.  The latter reaction cannot be directly measured, but is crucial input into the stellar models.   
Extending this program to heavier nuclei would quickly get impractical because of the need to introduce three- and four-body forces and multiple counter terms.

Another interesting neutrino-nucleus reaction is the coherent elastic neutrino scattering off nuclei, $\nu +A \rightarrow \nu +A$. This is a Standard Model process, but only recently has been observed \cite{Akimov:2017ade}. 

Neutrino-nucleus coherent elastic scattering differential cross section is given by \cite{Freedman:1977xn}
\begin{equation}
\label{coherent}
\frac{d \sigma}{dT} (E_{\nu},T) = \frac{G_F^2}{8 \pi} M \left[ 2 - \frac{2T}{T_{\rm max}} +  \left( \frac{T}{E_{\nu}} \right)^2 \right] Q_W^2 \left[ F(Q^2)\right]^2  
\end{equation}
where $E_{\nu}$ is the energy of the incoming neutrino, $M$ and $T$ are the mass and the recoil energy of the target nucleus, respectively.  $T_{\rm max}$ is the maximum value of $T$.  The weak charge of the nucleus, 
\begin{equation}
Q_W = N -(1-4 \sin^2\theta_W)Z,  \nonumber 
\end{equation}
primarily receives contributions from neutrons since $\sin^2 \theta_W \sim 1/4$. 
The form factor $F(Q^2)$, which is a function of the momentum transfer $Q$, 
corrects for contributions to scattering that are not completely coherent as $E_{\nu}$ gets large.  Contributions of the neutron density to this form factor is dominant since the proton density is again suppressed because of the smallness of the factor $4 \sin^2 \theta_W - 1$. Indeed this reaction was proposed as a tool to measure neutron densities inside nuclei 
\cite{Patton:2012jr,Patton:2013nwa}. It can also be useful in supernova detection \cite{Horowitz:2003cz}. Coherent elastic scattering of solar and atmospheric neutrinos is the background for the  experiments searching for particle dark matter by measuring the recoil of target nuclei after they are struck by dark matter particles. 

Integration of Eq. (\ref{coherent}) over nuclear recoil energies yields the total elastic cross section. If one ignores the nuclear form factor (i.e, $F(Q^2) =1$) this yields 
$\sigma \propto E_{\nu}^2$ as expected. However inclusion of nuclear structure effects reduces the cross section from this maximal value. Hence a careful calculation of the nuclear structure effects is important if one would like use this process as a probe to explore other physics such as the flux loss due to active-sterile neutrino mixing \cite{Anderson:2012pn}. 

One should mention that there are also subdominant contributions to the coherent elastic neutrino nucleus cross section such as those coming from non-zero neutrino magnetic moments. In a minimally extended Standard Model these contributions are expected to be finite, but very small. However new physics beyond the Standard Model may substantially increase them 
\cite{Balantekin:2013sda}. 

Most of the carbon in organic scintillators is in the form of $^{12}$C. Since the natural abundance of $^{13}$C is 1.07 \%, a sizable detector would already contain a substantial amount of this isotope.  SFO Hamiltonian, enhancing monopole terms of the matrix elements in the $p_{1/2}$ and $p_{3/2}$ orbitals, includes tensor components consistent with the general sign rule for the tensor-monopole terms \cite{Suzuki:2003,Otsuka:2005zz,Suzuki:2007zza}. A persistent problem for weak interactions in nuclei is the need to quench the axial-vector 
coupling strength $g_A$. Part of this quenching comes from the limited size of the model space and the effective interactions used. 
Calculations with this Hamiltonian reproduces the measured neutrino-$^{12}$C cross sections with a reduced quenching of $g_A$, as compared to the previous calculations \cite{Suzuki:2006qd}. These cross sections at the reactor energies  are calculated in Ref. \cite{Suzuki:2012aa}. It was found that a configuration space including up to $2\hbar \omega$ interactions with a small (five percent) quenching of the $g_A$ and spin g factor, this Hamiltonian considerably improves the cross sections as compared with the earlier treatments using Cohen-Kurath interactions \cite{Fukugita:1989wv}. 

\section{Reactor neutrino flux}

Short-baseline reactor neutrino experiments successfully measured the neutrino parameters they set out to measure, but they also identified a shape distortion in the 5-7 MeV range as well as a reduction from the predicted value of the flux \cite{An:2015nua}. 
This result and some of the other anomalies observed in neutrino experiments can be interpreted as mixing of  sterile neutrinos with active ones \cite{Abazajian:2012ys}. It was argued that there exists a discrepancy in reactor neutrino experiments between observed antineutrino fluxes near the reactor core and the predicted values \cite{Mention:2011rk}. 
This anomaly can be fitted with additional sterile neutrino states. 

Sterile neutrino explanation of the reactor flux discrepancy  is not a universally agreed conclusion \cite{Balantekin:2016vjt}. 
A careful analysis concludes that the corrections that lead to the reactor antineutrino anomaly are uncertain for the 30\% of the flux that arises from forbidden decays \cite{Hayes:2013wra}. 
Very recently the flux of neutrinos coming from the fissions of 
$^{235}$U and $^{239}$Pu in the cores of Daya Bay reactors were measured \cite{An:2017osx} and were found to be about 5\% less than predictions of the models 
\cite{Huber:2011wv,Mueller:2011nm}. 
Uncertainties in the subdominant corrections to beta-decay dominate the reactor neutrino spectra \cite{Hayes:2016qnu}, the resolution of which would require measuring fission products of many isotopes  \cite{Sonzogni:2017wxy}. 
For example three beta decays $^{92}$Rb, $^{96}$Y, and $^{142}$Cs contribute 43\% of the antineutrino flux emitted by nuclear reactors near 5.5 MeV. 
The latest measurement of these beta decays substantially modify the feedings of $^{142}$Ba from  $^{142}$Cs decays, increasing the discrepancy between the observed and the expected reactor antineutrino flux between 5 and 7 MeV \cite{Rasco:2016leq}. One way to estimate the reactor neutrino spectra is first to measure electron spectra from thermal fission products and convert that to neutrino spectra. In this method many fission products are measured together in a single experiment. It was pointed out \cite{Sonzogni:2017wxy} that including a shape correction of about +6\% MeV$^{-1}$ in conversion calculations fits the experimental Daya Bay spectrum better. 

The ultimate resolution of this issue from the neutrino side lies in further experiments as one needs to precisely measure any relative distortion of the $\bar{\nu}_e$ spectrum as a function of both energy and baseline.  PROSPECT, a precision oscillation and spectrum experiment, located at the High Flux Isotope Reactor (HFIR) at ORNL will measure the antineutrinos from a research reactor at a distance of less than 10m to resolve these questions \cite{Ashenfelter:2015uxt}. From the nuclear side one could envision measuring precise electron spectra of 50 or so fission products that can contribute. The shape factors for a least some of these can, in principle, be explored in rare ion facilities such as the Facility for Rare Ion Beams.

\section{Experimental Outlook}

Recent developments with experimental techniques made it possible to measure charge-exchange reactions with unprecedented precision. This development enables nuclear experimentalists to make a very precise determination of the Gamow-Teller strength distributions. For example the rate of the reaction $^{71}$Ga($\nu_e,e^-$) was recently deduced from the ($^3$He,$t$) charge-exchange reaction, 
leading to a slight change in the capture rate of the solar neutrinos coming from the pp reaction \cite{Frekers:2015wga}. 

Direct measurements of neutrino-nucleus cross sections are possible with intense neutrino sources. For relatively low energies, aside from nuclear power reactors, the list of such sources may include spallation neutron sources and beta beam facilities. In spallation neutron sources one can obtain a rather intense neutrino flux. Pulsed nature of this neutrino 
flux can then be used to eliminate much of the background \cite{Bolozdynya:2012xv}. Indeed such a facility was used to measure the coherent elastic neutrino-nucleus scattering 
\cite{Akimov:2017ade}.  Beta-beam facilities were proposed, but they are not currently under consideration. In such facilities beta-decay of boosted radioactive nuclei can be used to obtain an intense, collimated and pure neutrino beam. For low-energy neutrino-nucleus cross section measurements one can either use a low energy beta beam \cite{Volpe:2003fi} 
or utilize lower energy neutrinos at off-axis from a high-energy beta-beam \cite{Lazauskas:2007va}.

%uncomment the following lines to place a figure
%\begin{figure}[htb]
%\centerline{%
%\includegraphics[width=12.5cm]{Fig1}}
%\caption{Plot of ...}
%\label{Fig:F2H}
%\end{figure}

\vskip 0.3cm

This work was supported in part by the US National Science 
Foundation Grant No. PHY-1514695.


\begin{thebibliography}{99}


%\cite{An:2016srz}
\bibitem{An:2016srz} 
  F.~P.~An {\it et al.} [Daya Bay Collaboration],
  %``Improved Measurement of the Reactor Antineutrino Flux and Spectrum at Daya Bay,''
  Chin.\ Phys.\ C {\bf 41}, no. 1, 013002 (2017)
  doi:10.1088/1674-1137/41/1/013002
  [arXiv:1607.05378 [hep-ex]].
  %%CITATION = doi:10.1088/1674-1137/41/1/013002;%%

%\cite{An:2015nua}
\bibitem{An:2015nua} 
  F.~P.~An {\it et al.} [Daya Bay Collaboration],
  %``Measurement of the Reactor Antineutrino Flux and Spectrum at Daya Bay,''
  Phys.\ Rev.\ Lett.\  {\bf 116}, no. 6, 061801 (2016)
  Erratum: [Phys.\ Rev.\ Lett.\  {\bf 118}, no. 9, 099902 (2017)]
  doi:10.1103/PhysRevLett.116.061801, 10.1103/PhysRevLett.118.099902
  [arXiv:1508.04233 [hep-ex]].
  %%CITATION = doi:10.1103/PhysRevLett.116.061801, 10.1103/PhysRevLett.118.099902;%%

%\cite{Abe:2015rcp}
\bibitem{Abe:2015rcp} 
  Y.~Abe {\it et al.} [Double Chooz Collaboration],
  %``Measurement of ?$_{13}$ in Double Chooz using neutron captures on hydrogen with novel background rejection techniques,''
  JHEP {\bf 1601}, 163 (2016). 
  %%CITATION = doi:10.1007/JHEP01(2016)163;%%
  
%\cite{RENO:2015ksa}
\bibitem{RENO:2015ksa} 
  J.~H.~Choi {\it et al.} [RENO Collaboration],
  %``Observation of Energy and Baseline Dependent Reactor Antineutrino Disappearance in the RENO Experiment,''
  Phys.\ Rev.\ Lett.\  {\bf 116}, 211801 (2016). 
  %%CITATION = doi:10.1103/PhysRevLett.116.211801;%%  

%\cite{Alvarez-Ruso:2017oui}
\bibitem{Alvarez-Ruso:2017oui} 
  L.~Alvarez-Ruso {\it et al.},
  %``NuSTEC White Paper: Status and Challenges of Neutrino-Nucleus Scattering,''
  arXiv:1706.03621 [hep-ph].
  %%CITATION = ARXIV:1706.03621;%%




%\cite{Butler:1999sv}
\bibitem{Butler:1999sv} 
  M.~Butler and J.~W.~Chen,
  %``Elastic and inelastic neutrino deuteron scattering in effective field theory,''
  Nucl.\ Phys.\ A {\bf 675}, 575 (2000)
  doi:10.1016/S0375-9474(99)00682-X
  [nucl-th/9905059].
  %%CITATION = doi:10.1016/S0375-9474(99)00682-X;%%
  
 %\cite{Butler:2000zp}
\bibitem{Butler:2000zp} 
  M.~Butler, J.~W.~Chen and X.~Kong,
  %``Neutrino deuteron scattering in effective field theory at next-to-next-to-leading order,''
  Phys.\ Rev.\ C {\bf 63}, 035501 (2001)
  doi:10.1103/PhysRevC.63.035501
  [nucl-th/0008032].
  %%CITATION = doi:10.1103/PhysRevC.63.035501;%% 

%\cite{Butler:2002cw}
\bibitem{Butler:2002cw} 
  M.~Butler, J.~W.~Chen and P.~Vogel,
  %``Constraints on two-body axial currents from reactor anti-neutrino deuteron breakup reactions,''
  Phys.\ Lett.\ B {\bf 549}, 26 (2002)
  doi:10.1016/S0370-2693(02)02868-X
  [nucl-th/0206026].
  %%CITATION = doi:10.1016/S0370-2693(02)02868-X;%%

%\cite{Chen:2002pv}
\bibitem{Chen:2002pv} 
  J.~W.~Chen, K.~M.~Heeger and R.~G.~H.~Robertson,
  %``Constraining the leading weak axial two-body current by SNO and super-K,''
  Phys.\ Rev.\ C {\bf 67}, 025801 (2003)
  doi:10.1103/PhysRevC.67.025801
  [nucl-th/0210073].
  %%CITATION = doi:10.1103/PhysRevC.67.025801;%%

%\cite{Balantekin:2003ep}
\bibitem{Balantekin:2003ep} 
  A.~B.~Balantekin and H.~Yuksel,
  %``Constraints on axial two body currents from solar neutrino data,''
  Phys.\ Rev.\ C {\bf 68}, 055801 (2003)
  doi:10.1103/PhysRevC.68.055801
  [hep-ph/0307227].
  %%CITATION = doi:10.1103/PhysRevC.68.055801;%%

%\cite{Balantekin:2004zj}
\bibitem{Balantekin:2004zj} 
  A.~B.~Balantekin and H.~Yuksel,
  %``Neutrino physics and nuclear axial two-body interactions,''
  Int.\ J.\ Mod.\ Phys.\ E {\bf 14}, 39 (2005)
  doi:10.1142/S0218301305002758
  [nucl-th/0411025].
  %%CITATION = doi:10.1142/S0218301305002758;%%
  
%\cite{Shanahan:2017bgi}
\bibitem{Shanahan:2017bgi} 
  P.~E.~Shanahan {\it et al.},
  %``Isotensor Axial Polarizability and Lattice QCD Input for Nuclear Double-$\beta$ Decay Phenomenology,''
  Phys.\ Rev.\ Lett.\  {\bf 119}, no. 6, 062003 (2017)
  doi:10.1103/PhysRevLett.119.062003
  [arXiv:1701.03456 [hep-lat]].
  %%CITATION = doi:10.1103/PhysRevLett.119.062003;%%  
  
 %\cite{Akimov:2017ade}
\bibitem{Akimov:2017ade} 
  D.~Akimov {\it et al.} [COHERENT Collaboration],
  %``Observation of Coherent Elastic Neutrino-Nucleus Scattering,''
  Science {\bf 357}, no. 6356, 1123 (2017)
  doi:10.1126/science.aao0990
  [arXiv:1708.01294 [nucl-ex]].
  %%CITATION = doi:10.1126/science.aao0990;%% 
  
%\cite{Freedman:1977xn}
\bibitem{Freedman:1977xn} 
  D.~Z.~Freedman, D.~N.~Schramm and D.~L.~Tubbs,
  %``The Weak Neutral Current and Its Effects in Stellar Collapse,''
  Ann.\ Rev.\ Nucl.\ Part.\ Sci.\  {\bf 27}, 167 (1977).
  doi:10.1146/annurev.ns.27.120177.001123
  %%CITATION = doi:10.1146/annurev.ns.27.120177.001123;%%  
  
%\cite{Patton:2012jr}
\bibitem{Patton:2012jr} 
  K.~Patton, J.~Engel, G.~C.~McLaughlin and N.~Schunck,
  %``Neutrino-nucleus coherent scattering as a probe of neutron density distributions,''
  Phys.\ Rev.\ C {\bf 86}, 024612 (2012)
  doi:10.1103/PhysRevC.86.024612
  [arXiv:1207.0693 [nucl-th]].
  %%CITATION = doi:10.1103/PhysRevC.86.024612;%%  
  
%\cite{Patton:2013nwa}
\bibitem{Patton:2013nwa} 
  K.~M.~Patton, G.~C.~McLaughlin and K.~Scholberg,
  %``Prospects for using coherent elastic neutrino-nucleus scattering to measure the nuclear neutron form factor,''
  Int.\ J.\ Mod.\ Phys.\ E {\bf 22}, 1330013 (2013).
  doi:10.1142/S0218301313300130
  %%CITATION = doi:10.1142/S0218301313300130;%%  
  
 %\cite{Horowitz:2003cz}
\bibitem{Horowitz:2003cz} 
  C.~J.~Horowitz, K.~J.~Coakley and D.~N.~McKinsey,
  %``Supernova observation via neutrino - nucleus elastic scattering in the CLEAN detector,''
  Phys.\ Rev.\ D {\bf 68}, 023005 (2003)
  doi:10.1103/PhysRevD.68.023005
  [astro-ph/0302071].
  %%CITATION = doi:10.1103/PhysRevD.68.023005;%% 
 
 %\cite{Anderson:2012pn}
\bibitem{Anderson:2012pn} 
  A.~J.~Anderson, J.~M.~Conrad, E.~Figueroa-Feliciano, C.~Ignarra, G.~Karagiorgi, K.~Scholberg, M.~H.~Shaevitz and J.~Spitz,
  %``Measuring Active-to-Sterile Neutrino Oscillations with Neutral Current Coherent Neutrino-Nucleus Scattering,''
  Phys.\ Rev.\ D {\bf 86}, 013004 (2012)
  doi:10.1103/PhysRevD.86.013004
  [arXiv:1201.3805 [hep-ph]].
  %%CITATION = doi:10.1103/PhysRevD.86.013004;%%
 
  
 %\cite{Balantekin:2013sda}
\bibitem{Balantekin:2013sda} 
  A.~B.~Balantekin and N.~Vassh,
  %``Magnetic moments of active and sterile neutrinos,''
  Phys.\ Rev.\ D {\bf 89}, no. 7, 073013 (2014)
  doi:10.1103/PhysRevD.89.073013
  [arXiv:1312.6858 [hep-ph]].
  %%CITATION = doi:10.1103/PhysRevD.89.073013;%% 
  



\bibitem{Suzuki:2003}
T.~Suzuki, R.~Fujimoto and T.~Otsuka, 
 Phys.\ Rev.\ C {\bf 67}, 044302 (2003).
  
  
 %\cite{Otsuka:2005zz}
\bibitem{Otsuka:2005zz}
  T. Otsuka, T. Suzuki, R. Fujimoto, H. Grawe, and Y. Akaishi, 
  %``Evolution of Nuclear Shells due to the Tensor Force,''
  Phys.\ Rev.\ Lett.  {\bf 95}, 232502 (2005). 
  %%CITATION = PRLTA,95,232502;%%
  
%\cite{Suzuki:2007zza}
\bibitem{Suzuki:2007zza}
  T. Suzuki, S. Chiba, T. Yoshida, K. Higashiyama, M. Honma, T. Kajino, and T. Otsuka,  
  %``Advances in shell-model calculations and neutrino-induced reactions,''
  Prog.\ Part.\ Nucl.\ Phys.  {\bf 59}, 486 (2007). 
  %%CITATION = PPNPD,59,486;%%  
  
%\cite{Suzuki:2006qd}
\bibitem{Suzuki:2006qd}
  T. Suzuki, S. Chiba, T. Yoshida, T. Kajino, and T. Otsuka,  
  %``Neutrino nucleus reactions based on new shell model Hamiltonians,''
  Phys.\ Rev.\ C {\bf 74}, 034307(2006).
  [nucl-th/0608056].
  %%CITATION = NUCL-TH/0608056;%%  

%\cite{Suzuki:2012aa}
\bibitem{Suzuki:2012aa} 
  T.~Suzuki, A.~B.~Balantekin and T.~Kajino,
  %``Neutrino Capture on $^{13}$C,''
  Phys.\ Rev.\ C {\bf 86}, 015502 (2012)
  [arXiv:1204.4231 [nucl-th]].
  %%CITATION = ARXIV:1204.4231;%%

 
%\cite{Fukugita:1989wv}
\bibitem{Fukugita:1989wv} 
  M.~Fukugita, Y.~Kohyama, K.~Kubodera and T.~Kuramoto,
  %``REACTION CROSS-SECTIONS FOR neutrino C-13 ---> e- N-13 AND neutrino C-13 ---> neutrino-prime C*13 FOR LOW-ENERGY NEUTRINOS,''
  Phys.\ Rev.\ C {\bf 41}, 1359 (1990); 
  %%CITATION = PHRVA,C41,1359;%%
%\cite{Pourkaviani:1991pb}
%\bibitem{Pourkaviani:1991pb} 
  M.~Pourkaviani and S.~L.~Mintz,
  %``Neutrino reaction in C-13,''
  J.\ Phys.\ G G {\bf 17}, 1139 (1991); 
  %%CITATION = JPHGB,G17,1139;%%  
  %\cite{Mintz:2000xv}
%\bibitem{Mintz:2000xv} 
  S.~L.~Mintz,
  %``Inclusive neutrino reactions in C-13,''
  Nucl.\ Phys.\ A {\bf 672}, 503 (2000).
  %%CITATION = NUPHA,A672,503;%%

 
%\cite{Abazajian:2012ys}
\bibitem{Abazajian:2012ys} 
  K.~N.~Abazajian {\it et al.},
  %``Light Sterile Neutrinos: A White Paper,''
  arXiv:1204.5379 [hep-ph].
  %%CITATION = ARXIV:1204.5379;%%  
  
%\cite{Mention:2011rk}
\bibitem{Mention:2011rk} 
  G.~Mention, M.~Fechner, T.~Lasserre, T.~A.~Mueller, D.~Lhuillier, M.~Cribier and A.~Letourneau,
  %``The Reactor Antineutrino Anomaly,''
  Phys.\ Rev.\ D {\bf 83}, 073006 (2011)
  doi:10.1103/PhysRevD.83.073006
  [arXiv:1101.2755 [hep-ex]].
  %%CITATION = doi:10.1103/PhysRevD.83.073006;%%  
  
  
  
  
  
%\cite{Balantekin:2016vjt}
\bibitem{Balantekin:2016vjt} 
  A.~B.~Balantekin,
  %``Reactor antineutrinos and nuclear physics,''
  Eur.\ Phys.\ J.\ A {\bf 52}, no. 11, 341 (2016).
  doi:10.1140/epja/i2016-16341-5
  %%CITATION = doi:10.1140/epja/i2016-16341-5;%%  
  
%\cite{Hayes:2013wra}
\bibitem{Hayes:2013wra} 
  A.~C.~Hayes, J.~L.~Friar, G.~T.~Garvey, G.~Jungman and G.~Jonkmans,
  %``Systematic Uncertainties in the Analysis of the Reactor Neutrino Anomaly,''
  Phys.\ Rev.\ Lett.\  {\bf 112}, 202501 (2014)
  doi:10.1103/PhysRevLett.112.202501
  [arXiv:1309.4146 [nucl-th]].
  %%CITATION = doi:10.1103/PhysRevLett.112.202501;%%  
  
  
%\cite{An:2017osx}
\bibitem{An:2017osx} 
  F.~P.~An {\it et al.} [Daya Bay Collaboration],
  %``Evolution of the Reactor Antineutrino Flux and Spectrum at Daya Bay,''
  Phys.\ Rev.\ Lett.\  {\bf 118}, no. 25, 251801 (2017)
  doi:10.1103/PhysRevLett.118.251801
  [arXiv:1704.01082 [hep-ex]].
  %%CITATION = doi:10.1103/PhysRevLett.118.251801;%%
  
 %\cite{Huber:2011wv}
\bibitem{Huber:2011wv} 
  P.~Huber,
  %``On the determination of anti-neutrino spectra from nuclear reactors,''
  Phys.\ Rev.\ C {\bf 84}, 024617 (2011)
  Erratum: [Phys.\ Rev.\ C {\bf 85}, 029901 (2012)]
  doi:10.1103/PhysRevC.85.029901, 10.1103/PhysRevC.84.024617
  [arXiv:1106.0687 [hep-ph]].
  %%CITATION = doi:10.1103/PhysRevC.85.029901, 10.1103/PhysRevC.84.024617;%% 
  
 %\cite{Mueller:2011nm}
\bibitem{Mueller:2011nm} 
  T.~A.~Mueller {\it et al.},
  %``Improved Predictions of Reactor Antineutrino Spectra,''
  Phys.\ Rev.\ C {\bf 83}, 054615 (2011)
  doi:10.1103/PhysRevC.83.054615
  [arXiv:1101.2663 [hep-ex]].
  %%CITATION = doi:10.1103/PhysRevC.83.054615;%% 
  
  
 %\cite{Hayes:2016qnu}
\bibitem{Hayes:2016qnu} 
  A.~C.~Hayes and P.~Vogel,
  %``Reactor Neutrino Spectra,''
  Ann.\ Rev.\ Nucl.\ Part.\ Sci.\  {\bf 66}, 219 (2016)
  doi:10.1146/annurev-nucl-102115-044826
  [arXiv:1605.02047 [hep-ph]].
  %%CITATION = doi:10.1146/annurev-nucl-102115-044826;%% 
  
 %\cite{Sonzogni:2017wxy}
\bibitem{Sonzogni:2017wxy} 
  A.~A.~Sonzogni, E.~A.~McCutchan and A.~C.~Hayes,
  %``Dissecting Reactor Antineutrino Flux Calculations,''
  Phys.\ Rev.\ Lett.\  {\bf 119}, no. 11, 112501 (2017).
  doi:10.1103/PhysRevLett.119.112501
  %%CITATION = doi:10.1103/PhysRevLett.119.112501;%% 
  
 %\cite{Rasco:2016leq}
\bibitem{Rasco:2016leq} 
  B.~C.~Rasco {\it et al.},
  %``Decays of the Three Top Contributors to the Reactor $\overline ?_e$ High-Energy Spectrum, $^{92}$Rb , $^{96gs}$Y , and $^{142}$Cs , Studied with Total Absorption Spectroscopy,''
  Phys.\ Rev.\ Lett.\  {\bf 117}, no. 9, 092501 (2016).
  doi:10.1103/PhysRevLett.117.092501
  %%CITATION = doi:10.1103/PhysRevLett.117.092501;%%  
   
  
 %\cite{Ashenfelter:2015uxt}
\bibitem{Ashenfelter:2015uxt} 
  J.~Ashenfelter {\it et al.} [PROSPECT Collaboration],
  %``The PROSPECT Physics Program,''
  J.\ Phys.\ G {\bf 43}, no. 11, 113001 (2016)
  doi:10.1088/0954-3899/43/11/113001
  [arXiv:1512.02202 [physics.ins-det]].
  %%CITATION = doi:10.1088/0954-3899/43/11/113001;%% 
  
%\cite{Frekers:2015wga}
\bibitem{Frekers:2015wga} 
  D.~Frekers {\it et al.},
  %``Precision evaluation of the $^{71}$Ga($\nu_e,e^?$) solar neutrino capture rate from the ($^3$He,$t$) charge-exchange reaction,''
  Phys.\ Rev.\ C {\bf 91}, no. 3, 034608 (2015).
  doi:10.1103/PhysRevC.91.034608
  %%CITATION = doi:10.1103/PhysRevC.91.034608;%%  
  
%\cite{Freeman:2017bak}
\bibitem{Freeman:2017bak} 
  S.~J.~Freeman {\it et al.},
  %``An experimental study of the rearrangements of valence protons and neutrons amongst single-particle orbits during double {\beta} decay in 100Mo,''
  arXiv:1710.10817 [nucl-ex].
  %%CITATION = ARXIV:1710.10817;%%  
  
%\cite{Bolozdynya:2012xv}
\bibitem{Bolozdynya:2012xv} 
  A.~Bolozdynya {\it et al.},
  %``Opportunities for Neutrino Physics at the Spallation Neutron Source: A White Paper,''
  arXiv:1211.5199 [hep-ex].
  %%CITATION = ARXIV:1211.5199;%%  
  
%\cite{Volpe:2003fi}
\bibitem{Volpe:2003fi} 
  C.~Volpe,
  %``What about a beta beam facility for low-energy neutrinos?,''
  J.\ Phys.\ G {\bf 30}, L1 (2004)
  doi:10.1088/0954-3899/30/7/L01
  [hep-ph/0303222].
  %%CITATION = doi:10.1088/0954-3899/30/7/L01;%%  
  
 %\cite{Lazauskas:2007va}
\bibitem{Lazauskas:2007va} 
  R.~Lazauskas, A.~B.~Balantekin, J.~H.~De Jesus and C.~Volpe,
  %``Low-energy neutrinos at off-axis from a standard beta-beam,''
  Phys.\ Rev.\ D {\bf 76}, 053006 (2007)
  doi:10.1103/PhysRevD.76.053006
  [hep-ph/0703063 [HEP-PH]].
  %%CITATION = doi:10.1103/PhysRevD.76.053006;%% 
  
\end{thebibliography}
\end{document}